\documentclass[%
preprint,
 amsmath,amssymb,
 aps,
]{revtex4-1}

\usepackage{graphicx}
\usepackage{hyperref}

\usepackage{natbib}
\begin{document}
\title[Time Energy Relations]{A Class of Time-Energy Uncertainty Relations\\ for Time-dependent Hamiltonians}
\author{Tien D. Kieu}
\email{tien.d.kieu@gmail.com}
\affiliation{Centre for Quantum and Optical Science,
Swinburne University of Technology, Australia}
\begin{abstract}
A new class of time-energy uncertainty relations is directly derived from the Schr\"odinger equations for time-dependent Hamiltonians. Only the initial states and the Hamiltonians, but neither the instantaneous eigenstates nor the full time-dependent wave functions, which would demand a full solution for a time-dependent Hamiltonian, are required for our time-energy relations.  Explicit results are then presented for particular subcases of interest for time-independent Hamiltonians and also for time-varying Hamiltonians employed in adiabatic quantum  computation.

Some estimates of the lower bounds on computational time are given for general adiabatic quantum algorithms, with Grover's search as an illustration.  We particularly emphasise the role of required energy resources, besides the space and time complexity, for the physical process of (quantum) computation in general.
\end{abstract}
 \maketitle

\section{Introduction and overview of the results}
Similar to the position-momentum uncertainty relation (PMUR), the time-energy uncertainty relation (TEUR) is a property of the conjugacy of the Fourier transform variables time and frequency and of the quantum relationship between energy and frequency. But unlike the PMUR, the TEUR cannot be derived from any commutator relation, due to the lack of a well-defined time operator in quantum mechanics.  As a consequence, the interpretation of the TEUR is not as simple as that of the PMUR and has lead to serious misinterpretations~\cite{peres2006quantum}.  

The uncertainty relation between energy and time in nonrelativistic quantum mechanics was derived~\cite{Mandelshtam,Fleming} as the time it takes for a state with given energy spread to decay as measured through the rate of change of the expectation value of some observable.  Anandan and Aharonov~\cite{AnandanAharonov} have also applied geometric methods to quantum evolution to obtain certain TEUR for time-dependent Halmiltonians.
There are also other time energy relations expressing the least time it takes for the system under investigation to evolve into an orthogonal state in terms of the energy and energy spread of the system~\cite{margolus1998maximum,Brody, Deffner,Russell}, . See~\cite{Dodonov,DeffnerCampbell} for some recent reviews of the various TEURs and their derivations and interpretation as quantum speed limits.

In this paper we derive different sets of TEURs directly from the Schr\"odinger equations for time-dependent Hamiltonians in the general case.  Our employed method of derivation as well as the results are new and different to those in the existing literature.  It is important to note that only the initial states and the Halmiltionians, and neither the instantaneous eigenstates nor the full time-dependent wave functions, which would demand a full solution for a time-dependent Hamiltonian, are required for our time-energy relations.  This hallmark enables wider applicability and usefulness of our TEURs.

\subsection{Subcase of time-independent Hamiltonians}
As a special case for time-independent Hamiltonians, and as a {\em necessary but not sufficient} condition, is a lower bound on the time $\Delta t_\forall$ it takes for an initial state to evolve into any {\em arbitrary} state allowed by the Hamiltonian dynamics:
\begin{eqnarray}
{2\hbar} \le \Delta t_\forall \times{\sqrt{{\Delta E_0}^2 +(E_0)^2}}, \label{1}
 \end{eqnarray}
where $E_0$ and $\Delta E_0$ are, respectively the energy expectation value and the energy spread of the initial state. 

Also, as a {\em necessary but not sufficient} condition is a relation between the energy and the lower bound on the time $\Delta t_\perp$ it takes for an initial state to evolve into an {\em orthogonal} state allowed by the Hamiltonian dynamics:
\begin{eqnarray}
{\hbar\sqrt 2} \le \Delta t_\perp \times{{\Delta E_0}}. \label{2}
\end{eqnarray}

The necessary conditions above can also be expressed differently but equivalently as that the system cannot fully explore the whole Hilbert space (that is, cannot reach certain dynamically allowable state) or evolve into an orthogonal state from the initial state if the evolution time is {\em less than}, respectively, the characteristic times $T_\forall$ or $T_\perp$:
\begin{eqnarray}
T_\forall &\equiv& 2\hbar/\sqrt{{\Delta E_0}^2 +(E_0)^2}, \nonumber\\
T_\perp &\equiv& \hbar\sqrt{2}/{\Delta E_0}. \label{characteristic}
\end{eqnarray}

\subsection{Subcase of quantum adiabatic computation}
In addition, we also obtain some results for the adiabatic quantum computation AQC~\cite{Farhia,Farhib} with time-varying Hamiltonians.

AQC starts with the readily constructible ground state $|g_I\rangle$
of an initial Hamiltonian $H_I$ which is then {\em adiabatically}
extrapolated to the final Hamiltonian $H_P$ whose ground state
$|g\rangle$ encodes the desirable
solution of the problem and could be then obtained with
reasonably high probability.	The interpolation between $H_I$ and
$H_P$ is facilitated by a time-dependent Hamiltonian in the time
interval $0\le t \le T$,
\begin{eqnarray}
{\cal H}(t) &=& f(t/T)H_I + g(t/T)H_P,
\label{extrapolation}
\end{eqnarray}
either in a temporally linear manner (that is, $f(t/T) =
\left(1-{t}/{T}\right)$ and $g(t/T)={t}/{T} $); or otherwise with $f(0) = 1 = g(1)$ and $f(1) = 0 = g(0)$.  We also assume that both $f$ and $g$ are continuous, and $g$ is semi-positive for all $t\in[0,T]$.	Such a time
evolution is captured by the following Schr\"odinger equation:
\begin{eqnarray}
i\partial_t |\psi(t)\rangle &=& {\cal H}(t)\;|\psi(t)\rangle,
\label{AQC}\\
|\psi(0)\rangle &=& |g_I\rangle. \nonumber
\end{eqnarray}

We can set, without loss of generality, the initial ground state energy to zero to obtain the various {\em necessary} conditions, similar to the time-independent Hamiltonian above, for the computing time $T_{AQC}$ at the end of the computation:
\begin{eqnarray}
2\hbar &\le& T_{AQC\forall} \times \left(\int_0^1 g(\tau)d\tau\right)  \times\sqrt{\Delta_P E^2 +(E_P)^2}, \label{3} \\
\hbar\sqrt{2} &\le& T_{AQC\perp} \times \left(\int_0^1 g(\tau)d\tau \right) \times{\Delta_P E}, \label{4}
\end{eqnarray}
here $E_P$ and and $\Delta_P E$ respectively are the expectation energy and the energy spread of the {\em initial} state in terms of the {\em final} Hamiltonian:
\begin{eqnarray}
E_P &\equiv& \langle g_I | H_P | g_I \rangle,\nonumber\\
\Delta_P E &\equiv& {\sqrt{\langle g_I|H_P^2|g_I\rangle -\langle g_I|H_P|g_I\rangle^2}}.
\label{spread}
\end{eqnarray}
Similarly to the expressions~(\ref{characteristic}) above,  we also obtain 
the following AQC characteristic times:
\begin{eqnarray}
{\cal T}_{AQC\forall} &\equiv& \frac{2\hbar}{\left(\int_0^1 g(\tau)d\tau\right) \times\sqrt{{\Delta_P E}^2 +(E_P)^2}}, \label{AQClowerbound}\nonumber\\
{\cal T}_{AQC\perp} &\equiv& \frac{\hbar\sqrt 2}{\left(\int_0^1 g(\tau)d\tau\right) \times{\Delta_P E}^2}. \label{AQCcharacteristic}
\end{eqnarray}
That is, if the computation time is less than ${\cal T}_{AQC\forall}$ then there exists some state which is allowed by the dynamics but cannot be reached from the initial state.  And for evolution time less than ${\cal T}_{AQC\perp}$, the system cannot evolve to {\em any} state that is orthogonal to the initial state.  


The characteristic times in~(\ref{AQClowerbound}) are the lower bounds on the computing time; and as such, the more the energy and the more the spread of the {\em initial} state in energy with respect to the {\em final} Hamiltonian, the less the lower bound on computing time.  Note also that the inverses of these characteristic times are related to the measures of the interpolation function $g(t/T)$ of the AQC Hamiltonian~(\ref{extrapolation}).



Following are the derivations of the time-energy relations for the general cases and some of their consequences including those for AQC.  A lower bounds on the computing time for the adiabatic quantum Grover's search algorithms is also derived next.  The paper is then concluded with some remarks.

\section{Derivations for general time-dependent Hamiltonians}
From the Schr\"odinger equation with a time-dependent Hamiltonian:
\begin{eqnarray}
i\hbar\partial_t |\psi(t)\rangle &=& H(t)|\psi(t)\rangle, \label{Schroedinger}\\
|\psi(0)\rangle &=& |\phi_0\rangle,\nonumber
\end{eqnarray}
we also consider another state $|\phi(t)\rangle$ which satisfies a
closely related Schr\"odinger equation:
\begin{eqnarray}
i\hbar\partial_t |\phi(t)\rangle &=& \beta(t) {\bf 1}\;|\phi(t)\rangle,
\label{Schroedingerb}\\
|\phi(0)\rangle &=& |\phi_0\rangle, \nonumber
\end{eqnarray}
with arbitrary $c$-function $\beta(t)$.  This results in a phase ambiguity,
\begin{eqnarray}
|\phi(t)\rangle={\rm e}^{-i\int_0^t d\tau \beta(\tau)/\hbar}|\phi_0\rangle. \label{phase}
\end{eqnarray}
Here we consider the most general case of $\beta(t) \not = 0$ so that we could then exploit the freedom of choosing $\beta(t)$ in the following.

From the difference between Eqs.~(\ref{Schroedinger})
and~(\ref{Schroedingerb}), we have:
\begin{eqnarray}
\hbar\partial_t |\psi(t) - \phi(t)\rangle &=& -i H(t) |\psi(t) -
\phi(t)\rangle - i (H(t) -\beta(t))|\phi(t)\rangle.\label{diff}
\end{eqnarray}
We next consider:
\begin{eqnarray}
\hbar\partial_t \left\|\,|\psi(t)\rangle - |\phi(t)\rangle \right\|^2 &=&\hbar \partial_t \left(\langle \psi(t)- \phi(t)|\psi(t) - \phi(t)\rangle \right)\nonumber,\\
&=& \hbar
\left(\partial_t\langle\psi(t) -\phi(t)|\right)|\psi(t) - \phi(t)\rangle + 
\hbar\langle \psi(t)- \phi(t)|\left(\partial_t|\psi(t) -
\phi(t)\rangle\right), \nonumber\\
&=& 2\hbar\Re\,\langle \psi(t) - \phi(t)|
\left(\partial_t|\psi(t) -
\phi(t)\rangle\right).
\label{14}
\end{eqnarray}
Substituting~(\ref{diff}) for the time derivative on the rhs of the last expression and noting that $\langle \psi(t)-\phi(t)|H(t) |\psi(t)-\phi(t)\rangle$ is a real number due to the hermicity of $H(t)$, 
\begin{eqnarray}
{\rm rhs\;of~(\ref{14})}
&=& 2\Im\,\langle \psi(t) -
\phi(t)|(H(t) -\beta(t))|\phi(t)\rangle,
\nonumber\\
&\le&  2\left\|\,|\psi(t)
- \phi(t)\rangle\right\|\times\left\|(H(t)-\beta(t))|\phi(t)\rangle\right\|,\nonumber\\
&\le& 
2\left\|\,|\psi(t)
- \phi(t)\rangle\right\|\times\left\|(H(t) -\beta(t))|\phi_0\rangle\right\|,
\label{15}
\end{eqnarray}
where the first inequality is a result of the Schwarz inequality for the preceding rhs; and the second inequality is from~(\ref{phase}), where the entire time dependency of $|\phi(t)\rangle$ is residing in a phase factor. On the other hand, the lhs of~(\ref{14}) can also be rewritten as:
\begin{eqnarray}
\hbar\partial_t \left\|\,|\psi(t) - \phi(t)\rangle \right\|^2 &=&2\hbar\left\|\,|\psi(t) - \phi(t)\rangle \right\| \times \partial_t \left\|\,|\psi(t) - \phi(t)\rangle \right\| .\label{16}
\end{eqnarray}
Thus putting together~(\ref{15}) and~(\ref{16}) we then have:
\begin{eqnarray}
\hbar\partial_t \left\|\,|\psi(t) - \phi(t)\rangle \right\| &\le& 
\left\|(H(t) -\beta(t))|\phi_0\rangle\right\|.
\end{eqnarray}

Integrating the time
variable on both sides of the last expression from $0$ to $\Delta t$, and noting that $|\psi(0)\rangle = |\phi(0)\rangle$, we have the following inequality:
\begin{eqnarray}
\hbar{\left\|\,|\psi(\Delta t)\rangle - |\phi(\Delta t)\rangle
\right\|} &\le& \int^{\Delta t}_0 {\left\|(H(\tau) -\beta(\tau))|\phi_0\rangle\right\|}d\tau . \label{general}
\end{eqnarray}

The left hand side contains the distance $d(\Delta t, \beta(t))$ between the
final state of $|\psi(\Delta t)\rangle$ and the state $|\phi(\Delta t)\rangle$, which is the initial state up to a time-dependent phase factor depending on $\beta(t)$,
\begin{eqnarray}
d(\Delta t, \beta(t)) \equiv \left\|\,|\psi(\Delta t)\rangle - |\phi(\Delta t)\rangle \right\|. \nonumber
\end{eqnarray}
To refer back to the initial state $|\phi_0\rangle$, we have to set $\beta(t)=0$. With this,
the distance between the initial state and any dynamically allowable state in the corresponding Hilbert space is bounded by the least upper bound of two:
\begin{eqnarray}
d(\Delta t,\beta(t)=0) \le {2}. \nonumber
\end{eqnarray}
Denoting $\Delta t_\forall$ as the time when this bound is achieved, we then obtain from~(\ref{general}):
\begin{eqnarray}
\boxed{2\hbar \le \int^{\Delta t_\forall}_0{\left\|H(\tau)|\phi_0\rangle\right\|}d\tau .} \label{parallell}
\end{eqnarray}
This is the first general result.

We also have, from~(\ref{phase}),
\begin{eqnarray}
d(\Delta t, \beta(t)) &=& \sqrt{2-2\Re{\langle \psi(\Delta t)|\phi(\Delta t)\rangle}}, \nonumber \\
&=& \sqrt{2-2\Re({\rm e}^{-i\int_0^{\Delta t} d\tau \beta(\tau)/\hbar}{\langle \psi(\Delta t)|\phi_0\rangle)}}
\end{eqnarray}
If the initial state evolves into a state orthogonal to $|\phi_0\rangle$ at the time $\Delta t_\perp$, that is $\langle \psi(\Delta t_\perp)|\phi_0\rangle = 0$, then it is necessary that, {\em irrespective of} $\beta(t)$,
\begin{eqnarray}
d(\Delta t_\perp, \beta(t)) &=& \sqrt{2}.
\end{eqnarray}
Thus, we have from~(\ref{general}) a necessary condition at the time $\Delta t_\perp$:
\begin{eqnarray}
\boxed{\hbar\sqrt{2} \le \int^{\Delta t_\perp}_0{\left\|(H(\tau) -\beta(\tau))|\phi_0\rangle\right\|}d\tau .} \label{perpendicular}
\end{eqnarray}
This is the second general result.  Note that here we need not set $\beta(t) =0$ but will make use of this freedom of choice for $\beta(t)$ when we apply the general results~(\ref{parallell}) and~(\ref{perpendicular}) to the particular cases of time-independent Hamiltonians and time-dependent AQC in the below.

\section{For time-independent Hamiltonians}
For time-independent Hamiltonians we let $\beta$ be time-independent, and we also have:
\begin{eqnarray}
{\|(H-\beta)|\phi_0\rangle\|} &=&
{\sqrt{\langle \phi_0|(H-\beta)^2|\phi_0\rangle}}, \nonumber\\
&=& {\sqrt{\langle \phi_0|H^2|\phi_0\rangle
-2\beta \langle \phi_0|H|\phi_0\rangle + \beta^2}}, \nonumber\\
&=& {\sqrt{(\langle \phi_0|H^2|\phi_0\rangle -\langle
\phi_0|H|\phi_0\rangle^2) + (\beta - \langle
\phi_0|H|\phi_0\rangle)^2}}, \nonumber\\
&\equiv&\sqrt{{\Delta E_0}^2 +(E_0-\beta)^2},
\end{eqnarray}
where
\begin{eqnarray}
{\Delta E_0} &\equiv& {\sqrt{\langle \phi_0|H^2|\phi_0\rangle -\langle
\phi_0|H|\phi_0\rangle^2}}
\end{eqnarray}
is the energy spread of the initial state, and
\begin{eqnarray}
E_0 &\equiv& \langle \phi_0|H|\phi_0\rangle
\end{eqnarray}
is the energy expectation value of the initial state.

The general inequality~(\ref{parallell}) now reduces to:
\begin{eqnarray}
2\hbar &\le& {\Delta t_\forall} \times \sqrt{{\Delta E_0}^2 + E_0^2}. \label{26}
\end{eqnarray}
On the other hand, the inequality~(\ref{perpendicular}) becomes:
\begin{eqnarray}
\hbar\sqrt{2} &\le& {\Delta t_\perp} \times \sqrt{{\Delta E_0}^2 + (E_0-\beta)^2}.
\end{eqnarray}
We can now make use of the arbitrariness of $\beta$ to choose $\beta = E_0$ in order to minimise the rhs of the last expression, and thus tighten the inequality, to arrive at:
\begin{eqnarray}
\hbar\sqrt{2} &\le& {\Delta t_\perp} \times {{\Delta E_0}}. \label{28}
\end{eqnarray}
Expressions~(\ref{26}) and~(\ref{28}) are the advertised results~(\ref{1}) and~(\ref{2}).

Our characteristic times~(\ref{characteristic}) are of the same order of magnitudes, under corresponding conditions, as the text-book ETUR bound~\cite{peres2006quantum} and Margolus-Levitin's bound~\cite{margolus1998maximum}.
Expression~(\ref{28}) contains the numerical factor $\sqrt{2}$, which is mathematically slightly weaker than the factor of $\pi/2$ in the canonical result $\pi\hbar/2\lesssim< \Delta t\Delta E$.

We next derive another consequence of~(\ref{general}) for time-independent Hamiltonians.  With  $P(t) = \left|\langle \psi(t)|\phi_0\rangle\right|^2$ is the probability at time $t$ that the system remains in its initial state, we then have:
\begin{eqnarray}
d^2(t) &=& 2 - 2\Re\langle \psi(t)|\phi_0\rangle, \nonumber\\
&\ge& 2\left(1- \left| \langle \psi(t)|\phi_0\rangle\right|\right), \nonumber\\
&\ge& 2\left(1- \sqrt {P(t)}\right), \nonumber
\end{eqnarray}
from which:
\begin{eqnarray}
P(t) &\ge& \left( 1-\frac{1}{2}d^2(t)\right)^2.
\label{probability}
\end{eqnarray}
Now combining this with~(\ref{general}) with $\beta = E_0$ leads us to the result:
\begin{eqnarray}
P(t) &\ge& \left( 1-\frac{({\Delta E_0})^2}{2\hbar^2}t^2\right)^2.
\label{decay1}
\end{eqnarray}
In particular, for small $t\Delta E_0\ll \hbar$, the decay probability is slower than an exponential decay in $t$:
\begin{eqnarray}
P(t) &\gtrsim& {\rm \exp}\left\{ -\frac{({\Delta E_0})^2}{\hbar^2}t^2\right\}. \label{decay}
\end{eqnarray}

It is not difficult to apply the above arguments to a composite system and show that quantum entanglement between the system components will speed up the decay probability as compared to system with independent, non-entangled components.  Left as an open question but it would be interesting to establish a relationship between the degree of entanglement and that of the speeding up.


\section{For Adiabatic Quantum Computation}
To apply the general result~(\ref{general}) to AQC we replace $H(t)$ with the AQC Hamiltonian~(\ref{extrapolation}), and $|\phi_0\rangle$ with the ground state $|g_I\rangle$ of the initial Hamiltonian $H_I$, where without loss of generality we can let $H_I|g_I\rangle = 0$.  We also let $\beta(t) = \beta_0 g(t/T)$ where $\beta_0$ is time-independent.  We then have for the rhs of~(\ref{general}):
\begin{eqnarray}
\int^{\Delta t}_0 \left\| ({\cal H}(\tau) -\beta(\tau))|\phi_0\rangle \right\|d\tau 
&=& \left(\int_0^{\Delta t} g(\tau/T) d\tau \right)\times {\|(H_P-\beta_0)|g_I\rangle\|},
\end{eqnarray}
of which
\begin{eqnarray}
{\|(H_P-\beta_0)|g_I\rangle\|} &=&
{\sqrt{\langle g_I|(H_P-\beta_0)^2|g_I\rangle}}, \nonumber\\
&=& {\sqrt{\langle g_I|H_P^2|g_I\rangle
-2\beta_0 \langle g_I|H_P|g_I\rangle + \beta_0^2}}, \nonumber\\
&=& {\sqrt{(\langle g_I|H_P^2|g_I\rangle -\langle
g_I|H_P|g_I\rangle^2) + (\langle
g_I|H_P|g_I\rangle--\beta_0)^2}}, \nonumber\\
&\equiv& \sqrt{{\Delta_P E}^2 + (E_P-\beta_0 )^2}. 
\end{eqnarray}

The inequalities~(\ref{parallell}) and~(\ref{perpendicular}) now, at the end of the computation, reduce to:
\begin{eqnarray}
2\hbar &\le& {T_{AQC\forall}} \left(\int_0^1g(\tau)d\tau\right) \times \sqrt{{\Delta_P E}^2 +E_P^2},\\
\hbar\sqrt{2} &\le& {T_{AQC\perp}}\left(\int_0^1g(\tau)d\tau \right)\times {{\Delta_P E}}. \label{36}
\end{eqnarray}
These are the advertised results~(\ref{3}) and~(\ref{4}). Note that we have made the judicious choice of $\beta_0 = E_P$ to arrive at~(\ref{36}).

These inequalities incorporate the initial
ground state $|g_I\rangle$ and the spectrum of the final Hamiltonian in the forms of the energy expectation and the energy spread of the initial state in terms of the final energy $\Delta_P E$. The manner of the time extrapolation, which dictates the temporal flow of energy eigenstates of ${\cal H}(t)$ in~(\ref{extrapolation}), is further reflected in $g(\tau)$.

Note that corresponding results can also be derived for a
generalised AQC~\cite{
Wei, Boulatov2005}, where an extra
term of the form $h(t) H_E$ (with $h(0)=h(T)=0$) is added
to the Hamiltonian~(\ref{extrapolation}) to realise some further freedom in the adiabatic paths.

Similar to~(\ref{decay1}) we can also derive a lower bound on the decay probability $P_{AQC}(t) = |\langle \psi(t)|g_I\rangle|^2$ for AQC:
\begin{eqnarray}
P_{AQC}(t) &\ge& \left( 1-\frac{(\Delta E_P)^2}{2\hbar^2}\left(\int_0^t g(\tau/T)d\tau\right)^2\right)^2,
\end{eqnarray}
where $T$ is the end time of the computation at which ${\cal H}(T) = H_P$
.
  
\section{Quantum Speed Limit for Adiabatic Grover's Search}
We present below an application of our new TEURs~(\ref{36}) in an estimation of some lower bounds on the  computational time for Grover's unstructured search algorithms in AQC.  See~\cite{Lychkovskiy} for a different approach for lower bounds using other uncertainty relations where the full time dependence of the wave functions is required, in contrast to our results above.  We also emphasise here
the need for energy resources, not only in quantum computation but also in any physical computation, as an essential component for computational complexity, besides the usual resources of memory space and computing time.  

We first consider a quantum  algorithm~\cite{Roland2002, Wei} to locate the state $|m\rangle$ in an unsorted database set of normalised orthogonal states $\{|i\rangle, i = 1, \ldots, M\}$.  It is known that this algorithm has a computational complexity of ${\cal O}(\sqrt M)$ as that of Grover's algorithm~\cite{Grover}, a quadratic improvement on classical search.

For an AQC algorithm, we start with an initial state $|\phi_0\rangle$ that is a uniform superposition of all the states in the given search set,
\begin{eqnarray}
|\phi_0\rangle &\equiv& \sum_{i=1}^M c_i|i\rangle. \label{initial}
\end{eqnarray}
This state is the ground state of the initial Hamiltonian $H_I$ in~(\ref{extrapolation}),
\begin{eqnarray}
H_I &=& {\bf 1} - |\phi_0\rangle \langle \phi_0|.
\end{eqnarray}
The target Hamiltonian $H_P$ is then designed to have the solution state $|m\rangle$ as the ground state,
\begin{eqnarray}
H_P &=& {\bf 1} - |m\rangle \langle m|.
\end{eqnarray}
The AQC is performed in the usual manner with a time-dependent Hamiltonian ${\cal H}_G (t)$ in the time interval $t\in[0,T]$ according to~(\ref{extrapolation}),
\begin{eqnarray}
{\cal H}_G (t) &=& f(t/T)H_I + g(t/T)H_P. \label{H_G}
\end{eqnarray}

The energy expectation and energy spread of the target Hamiltonian as measured in the initial state $|\phi_0\rangle$ are, respectively,
\begin{eqnarray}
\langle \phi_0 | H_P |\phi_0 \rangle &=& 1 - |c_m|^2, \label{energy}
\end{eqnarray}
and
\begin{eqnarray}
\Delta_P E &\equiv& \sqrt{\langle \phi_0 | H_P^2 |\phi_0 \rangle - \langle \phi_0 | H_P |\phi_0 \rangle^2}, \nonumber\\
&=& \sqrt{|c_m|^2 - |c_m|^4}. \label{delta}
\end{eqnarray}

According to the time-energy uncertainty~(\ref{AQCcharacteristic}), the time estimate $T_\perp^{\rm search}$,
is the lower limit below which the initial state $|\phi_0\rangle$ {\em cannot} evolve into an orthogonal state under the dynamics governed by ${\cal H}_G(t)$ in~(\ref{H_G}).  This time limit should be of the same order of magnitude as the {\em best} AQC computation time, as estimated according to the quantum adiabatic theorem, to obtain the target state $|m\rangle$ with reasonable probability.

For the case of the initial state is a {\em uniform} superposition of all the states, that is,
\begin{eqnarray}
c_i &=& \frac{1}{\sqrt M}, \forall i = 1, \ldots, M,
\end{eqnarray}
we then have from~(\ref{AQCcharacteristic})
\begin{eqnarray}
T_\perp^{\rm search} &\stackrel{M\gg 1}{\sim}& {\cal O} \left(\frac{\sqrt M}{\int_0^1 g(\tau) d\tau}\right). \label{A8}
\end{eqnarray}
This time estimate, with $g$ for which $ \int_0^1 g(\tau) d\tau\sim {\cal O}(1)$, is indeed of the same order of magnitude as the time complexity ${\cal O}(\sqrt[]{M})$ for the AQC~\cite{Roland2002} as normally obtained from the energy gap of the two lowest eigenvalues in the spectral flow of ${\cal H}_G(t)$ according to the quantum adiabatic theorem.

In contrast to those derived from the quantum adiabatic theorem, the time estimate $T_\perp^{\rm search}$ here depends only on the extrapolating function $g$, the initial state and the target Hamiltonian.  Our lower bound estimate, furthermore for this particular algorithm,  is independent of all other amplitudes $c_i$ for $i\not= m$. It depends only on the coefficient $c_m$ of the target state in the superposition~(\ref{initial}).  We thus could improve on the time ${\cal O}(\sqrt{M})$ if we have some information that leads to higher priori probability for the target state $|m\rangle$, such that $|c_m| > 1/\sqrt{M}$.  

In addition to that, we could also exploit the extra degree of freedom of the extrapolating function $g$ to reduce the time estimate~(\ref{A8}).  For example, with the choice
\begin{eqnarray}
g(\tau) &\longrightarrow& \sqrt{M}g(\tau) \label{A10}
\end{eqnarray}
substituting in~(\ref{A8}) we could have reduced the lower time limit $T_\perp^{\rm search} \sim {\cal O}(1)$!  This choice and its computation time have also been confirmed in~\cite{Wei} by different means.

As another example, the authors of~\cite{0204044} employ a different function $g(\tau)$ but which also grows with $\sqrt M$,
\begin{eqnarray}
g(\tau) &\longrightarrow& \tau + \sqrt{M}\tau(1-\tau) \label{A11}.
\end{eqnarray}
This once again reduces the computation time to ${\cal O}(1)$, also in agreement with~(\ref{A8}) whence
\begin{eqnarray}
\int^1_0 g(\tau) d\tau &\longrightarrow& 1/6 + \sqrt{M}/2.
\end{eqnarray}

All of the above reductions for $T_\perp^{\rm search}$ match, in orders of magnitude, the time complexities derived in~\cite{0204044, Wei} directly from a consideration of spectral-flow energy gap according to the quantum adiabatic theorem.  This agreement is remarkable, as our results above are not derived directly from the quantum adiabatic theorem but from a general consideration of time-energy uncertainty relation for time-dependent Hamiltonians.

Such an agreement, however, is not unexpected.  It should be reminded again here that our time measure $T_\perp^{\rm search}$ is a necessary lower limit in the sense that if the computation time is less than that then the initial state cannot evolve into an orthogonal state.  But longer computation time, $T > T_\perp^{\rm search}$, is {\em not} a sufficient condition; for sufficiency we would need to involve the quantum adiabatic theorem.  However, the simply calculated time measure $T_\perp^{\rm search}$ does agree in order of magnitudes with the estimate of the computational time more comprehensively derived.  This agreement of our results and those in~\cite{Wei} demonstrates that these necessary lower limits can in fact be saturated in this case by judicious choice of the extrapolating functions $f$ and $g$~\cite{Roland2002, Wei}.

More importantly, we want to point out and emphasise here that although we may be able to reduce the time complexity to ${\cal O}(1)$, as with the choice of~(\ref{A10})or~(\ref{A11}), we need to consider also the energy resource required for the computation.  The choice of~(\ref{A10}) can, in fact, only be had at the cost of an increase in the energy required:
\begin{eqnarray}
\max_{0\le t \le T} \langle \phi_0| {\cal H}_G(t) |\phi_0\rangle &\to& {\cal O}(\sqrt M).
\end{eqnarray}
That is, a reduction in time complexity (from ${\cal O}(\sqrt M)$ to ${\cal O}(1)$) incurs and is balanced by an increase in the cost in energy resource (from ${\cal O}(1)$ to ${\cal O}(\sqrt M)$).  That is, in general the computational complexity of the AQC~(\ref{H_G}) is of the order ${\cal O}(\sqrt M)$, taking into account both the energy and time resources.

One could go to the extreme and implement the choice 
\begin{eqnarray}
g(\tau) &\longrightarrow& {\rm e}^Mg(\tau) \label{A12}
\end{eqnarray}
in order to have an exponentially {\em decreasing} computation time with $M$.  But of course we then at the same time have to pay exponentially for the energy required for the physical AQC computation,
\begin{eqnarray}
\max_{0\le t \le T} \langle \phi_0| {\cal H}_G(t) |\phi_0\rangle &\to& {\cal O}({\rm e}^M).
\end{eqnarray}

{\em The message here is that in considering the computational complexity in general we need to consider also the energy resources in addition to the usually considered space and time complexity.}  

See~\cite{0110020, 0204087, 0205048, 0208135, 0309201} for other AQC Hamiltonians for this search problem.  For these and more general time-dependent Hamiltonians which do not assume the particular form of~(\ref{extrapolation}), we still could estimate some lower bounds on the computational time through the general uncertainty relations~(\ref{parallell}) and~(\ref{perpendicular}).

\section{Concluding Remarks}
We derive a new class of time-energy uncertainty relations directly from the Sch\"odinger equationS for the general case of time-dependent Hamiltonians.  The results are then applied to the estimation of some lower bounds on computational time in quantum adiabatic computation (AQC).

Our general results for time-dependent Hamiltonians in~(\ref{parallell}) and~(\ref{perpendicular}) establish some fundamental upper bounds on the speed of quantum evolution. Such quantum speed limits are normally dependent on the choice of a measure of distinguishability of quantum states, but it has also been shown that they could be qualitatively governed by the norm of the generator of quantum dynamics~\cite{Deffner_2017}. 

However, our new results herein involve only the {\em initial} states and do not require the generally intractable solutions of the {\em full} dynamics and/or of the instantaneous energy eigenstates, as demanded in all other quantum speed limits derived elsewhere in the literature thus far.  Direct comparison of our bounds to others' would thus require solving for the full dynamics and/or the eigenstate problem of time-dependent Hamiltonians.  However, we have a clear and significant advantage in that our bounds are simpler and consequently have much wider applicability in general.  Such advantage has been demonstrated in the particular but important case of AQC.  

Our characteristic lower bounds on computing time for AQC depend only on the {\em initial} state of the computation, its energy expectation and also its energy spread as measured in terms of the {\em final} (target) Hamiltonian.  These estimates are not explicitly but only implicitly dependent on the instantaneous energy gaps of the time-dependent Hamiltonian at intermediate times.  As a result, the estimates have no bearing on the probability for obtaining the ground state of the final target Hamiltonian.  Nevertheless, in the case of adiabatic quantum algorithms for Grover's search, it is confirmed that such lower bounds are indeed saturated in the same order of magnitude as the computing time evaluated elsewhere from the quantum adiabatic theorem.

We emphasise that in considering the computational complexity of a quantum algorithm one would need to consider also the energy resources in addition to the usually considered space and time complexity.  This feature has also been illustrated in some adiabatic algorithms for the problems of travelling salesman and integer factorisation~\cite{Kieu_TSP, Kieu_factoring}.

\section*{Acknowledgement}
The author wishes to thank Adolfo del Campo, Peter Hannaford and the referees for helpful suggestions.  

\bibliography{ETUR}
\end{document}